\documentclass{pas}

\usepackage{multirow}
\usepackage{amsmath}

\begin{document}

\lefttitle{Publications of the Astronomical Society of Australia}
\righttitle{B. Amend \textit{et al.}}

\jnlPage{1}{15}
\jnlDoiYr{2026}
\doival{10.1017/pasa.xxxx.xx}

\articletitt{Research Paper}

\title{The interaction phase of engine-driven explosions and high-energy winds}

\author{
\gn{Benjamin} \sn{Amend}$^{1}$,
\gn{Christopher} \sn{Lagomarsino}$^{1}$,
\gn{Eric R.} \sn{Coughlin}$^{1}$ and
\gn{Jonathan} \sn{Zrake}$^{2}$
}

\affil{
$^1$Department of Physics, Syracuse University, Syracuse, NY 13210, USA\\
$^2$Department of Physics and Astronomy, Clemson University, Clemson, SC 29634, USA
}

\corresp{Benjamin Amend, Email: bjamend@syr.edu}

\citeauth{Amend B., Lagomarsino C., Coughlin E. R. and Zrake J. The interaction phase of engine-driven explosions and high-energy winds. {\it Publications of the Astronomical Society of Australia} {\bf 00}, 1--15. https://doi.org/xx.xxxx/xxxxx}

\history{(Received xx xx xxxx; revised xx xx xxxx; accepted xx xx xxxx)}

\begin{abstract}
Wide-angle outflows, or winds, are associated with a broad range of astrophysical systems, including protostars, massive stars, X-ray binaries, tidal disruption events (TDEs), luminous fast blue optical transients (LFBOTs), and starburst galaxies. When these winds first ``turn on," they inflate a ``bubble" into their surroundings, bounded by two shocks and a contact discontinuity, and evolve through distinct adiabatic phases prior to the onset of significant radiative cooling. For sufficiently overdense ejecta, the flow quickly relaxes into an interaction-dominated similarity state at early times and later enters an energy-conserving regime. We present a systematic study of these phases for adiabatic winds expanding into power-law density profiles $\rho \propto r^{-n}$ with $0 \leq n \leq 2$. Using analytic scalings together with one-dimensional shock-capturing hydrodynamic simulations, we quantify both the relaxation timescales and the accuracy with which the corresponding similarity solutions reproduce the fluid velocity, density, and pressure throughout the shocked bubble. We show that the interaction solutions are attained within only a few dynamical times and remain valid until the reverse-shocked shell is no longer thin relative to the forward-shocked shell, corresponding in practice to an instantaneous overdensity of order unity. For $n < 2$, the flow subsequently converges to the generalized energy-conserving scaling $R_s \propto t^{3/(5-n)}$, while the special case $n=2$ exhibits a single persistent similarity state. We discuss the durations and implications of these phases for stellar and galactic outflows, TDEs, and LFBOTs.
\end{abstract}


\maketitle

\section{Introduction}

Winds are a common and dynamically important feature of many astrophysical systems, including OB-type and protostellar winds \citep{cak75, 1977ApJ...218..377W, 2001A&A...369..574V}, Wolf-Rayet (WR) stars \citep{1972SvA....15..708A, 2007ARA&A..45..177C}, white dwarfs \citep{1982ApJ...260..716C}, stellar clusters \citep{1985Natur.317...44C}, active galactic nuclei \citep{2015ARA&A..53..115K, 1995ApJ...451..498M}, and starburst galaxies \citep[e.g.][]{2016MNRAS.455.1830T}. Related hydrodynamic structures may also arise in some tidal disruption events (TDEs) \citep{1988Natur.333..523R, 2021ARA&A..59...21G}, where the tidal destruction of a star by a supermassive black hole leads to rapid fallback of stellar debris and the formation of an often super-Eddington accretion flow \citep{1989ApJ...346L..13E, 2018MNRAS.478.3016W}. That accretion flow can in some cases launch a relativistic jet or disk wind \citep[e.g.][]{2011MNRAS.415..168S}. While only a small number of TDEs show clear relativistic jets or spectroscopically confirmed outflows, many now exhibit late-time radio emission consistent with an outflow interacting with the circumnuclear medium \citep{2011MNRAS.416.2102G,2021NatAs...5..491H,2024ApJ...971..185C,2024ApJ...971...49M}. The shocks driven by that interaction can produce a distinct observable flare, separate from the thermal optical, ultraviolet, and X-ray emission associated more directly with the accretion episode itself. Similar considerations may apply to at least some luminous fast blue optical transients (LFBOTs), for which central-engine activity has been proposed in a number of events and whose radio/X-ray observations likely imply fast ejecta interacting with dense circumstellar material \citep{Margutti_2019, 2019ApJ...871...73H, 2020ApJ...895...49H, 2020ApJ...895L..23C, 2022ApJ...934..104Y, 2026arXiv260103337P}.

In each case, the wind inflates a hot, pressurized ``bubble'' that is bounded by both a forward and reverse shock, where the former sweeps up and heats the ambient material and the latter (also called a ``wind termination shock") decelerates and pressurizes the otherwise-cold wind material \citep{cmw75}. These wind-driven bubbles play a central role in regulating feedback in the interstellar medium (ISM) \citep{1977ApJ...218..377W, 2013ApJ...776....1K, 2016MNRAS.456..710F, 2020ApJ...894L..24F}, and accurately characterizing their hydrodynamic evolution is therefore essential for understanding problems ranging from stellar feedback to large-scale galactic outflows.

When the wind ``turns on" and initially impacts its surroundings, radiative losses are dynamically insignificant and the system evolves adiabatically. The shocked regions then expand and cool while the dynamical timescales increase and radiative losses become more significant. As outlined in \cite{cmw75} and \cite{1977ApJ...218..377W}, radiative cooling is initially only important for the forward-shocked medium, which compresses into a thin and dense shell while the reverse-shocked interior remains hot and continues to drive the expansion through $p-dV$ work. As the system continues to evolve and the interior pressure declines due to continued expansion and cooling, radiative losses become important for the reverse-shocked wind as well, and the entire system ultimately approaches a momentum-conserving ``snowplow" phase \citep{1988RvMP...60....1O}.

In the original analyses of \cite{1977ApJ...218..377W} and \cite{1972SvA....15..708A}, motivated primarily by stellar feedback into a uniform ISM, the early adiabatic phase was found to be comparatively short-lived ($\sim 2\times10^3\,$yr) and therefore of limited physical significance. This phase corresponds to the classic energy-conserving stage, in which the shocked wind has relaxed to an approximately uniform-pressure interior, the reverse shock lies well interior to the contact discontinuity, the shocked ejecta density is well below that of the shocked ambient medium, and the flow is therefore well approximated as isobaric. In such environments, the dense swept-up shell cools efficiently, rapidly driving the flow into the subsequent radiative phases. However, for decreasing ambient medium density profiles ($\rho \propto r^{-n}$ where $n>0$), reduced post-shock densities are expected to increase the cooling time relative to the expansion time, allowing the adiabatic phase to persist substantially longer. Such scenarios could arise, for instance, when a newly-launched, faster stellar wind from a later evolutionary phase overtakes and sweeps up the slower, denser outflow from an earlier phase \citep[e.g.][]{1996A&A...316..133G}, when a galactic outflow propagates through a gravitationally- or hydrostatically-stratified halo or circumgalactic medium (CGM) \citep{1988ApJ...324..776M}, or when an outflow from a tidally disrupted star interacts with the circumnuclear medium.

Furthermore, \cite{2024ApJ...975L..14C} recently identified additional self-similar solutions that characterize the flow at even earlier times, before the onset of the aforementioned energy-conserving phase. At this stage, the freely expanding wind still interacts more directly with the ambient medium: the ratio $f$ of the wind density to the ambient density is high, the reverse shock is not well separated from the contact discontinuity, and the shocked ejecta is not well-approximated as isobaric. As with the later adiabatic stage, steeper ambient density gradients prolong this early regime by delaying the attainment of this smaller overdensity ratio (e.g., see the figures and analysis in the following sections).

Taken together, these considerations suggest that wind-driven systems evolve through multiple distinct similarity regimes whose durations and relative dynamical importance depend sensitively on the ambient medium density profile and initial density contrast. In lower-energy systems, this initial density contrast may be vanishingly small, so that the shocked wind rapidly becomes underdense relative to the shocked ambient medium, and the interior flow reaches the approximately isobaric structure of the energy-conserving phase with little or no extended interaction regime. By contrast, in higher-energy outflows, such as those associated with WRs, TDEs, LFBOTs, etc., the initial overdensity ratio can be extremely high, allowing for a more prolonged interaction-dominated phase as discussed in detail in Section~\ref{subsec:observational-implications}.

While the interaction solutions and the later energy-conserving solutions are each well understood in their respective limits, the transition between these regimes (and the accuracy with which they describe the full hydrodynamic evolution) has not been systematically explored, particularly for stratified ambient media with $\rho \propto r^{-n}$. In this work, we present a unified study of the adiabatic evolution of wind-driven bubbles expanding into power-law density profiles, comparing analytic and semi-analytic scalings to one-dimensional hydrodynamic simulations. Because we are interested in applications to high-energy systems, we largely consider $f \gg 1$. We quantify the ``relaxation timescales'' toward each similarity state and assess the accuracy of the solutions during transitions between self-similar regimes in different ambient density profiles.

This paper is structured as follows. In Section~\ref{sec:basic-considerations}, we derive scalings for relevant timescales and shock positions in both similarity regimes. We use Section~\ref{sec:numerical-methods} to describe the novel shock-capturing numerical scheme we implement to resolve the narrow shocked region of interest, and we state our simulation parameters and initial conditions. We compare our hydrodynamic simulations to the relevant self-similar solutions in their respective limits in Section~\ref{sec:results}, and interpret our results in the broader context of astrophysical winds in Section~\ref{sec:discussion}.
We conclude with a summary of our findings in Section~\ref{sec:conclusions}.

\section{Basic considerations}
\label{sec:basic-considerations}

We consider a time-steady, spherically symmetric wind launched at radius $R_{w,0}$ with constant mechanical luminosity $L_w=\frac{1}{2}\dot{M}v_w^2$, where $v_w$ is the wind velocity and $\dot{M}$ is the wind mass flux. The wind density is then $\rho_w(r) = \rho_{w,0}(r/R_{w,0})^{-2}$, where $\rho_{w,0} = \dot{M}/(4\pi v_w R_{w,0}^2)$. This wind expands into an ambient medium with density $\rho_a(r) = \rho_{a,0}(r/R_{w,0})^{-n}$, where $\rho_{a,0}$ is a constant. The collision of the wind and the ambient medium produces a double-shock structure: a reverse shock (or wind termination shock) at radius $R_{\mathrm{r}}(t)$, a forward shock at $R_{\mathrm{s}}(t)$, and a contact discontinuity at $R_{\mathrm{c}}(t)$ that separates the shocked wind from the shocked ambient gas. At the earliest times, when the wind density can greatly exceed that of the ambient medium, the dynamics are governed by direct ram-pressure interaction and rapidly approach the interaction-dominated, or ``coasting," self-similar regime described by \cite{2024ApJ...975L..14C}. Once the ambient and ejecta densities become comparable, it is expected that the system transitions to an energy-conserving expansion described by the first stage of the solutions presented in \cite{1977ApJ...218..377W, 1988RvMP...60....1O, 1992ApJ...388..103K}.

\subsection{Interaction phase}
\label{subsec:interaction-phase}\

At sufficiently early times, the freely expanding wind can be much denser than the ambient medium, $\rho_w \gg \rho_a$, and the shocked wind speed is comparable to the free-wind speed, $v_{\rm w}$. The forward shock propagates faster than the wind speed, such that the relative thickness of the forward-shocked shell,
\begin{equation}
    \Delta_{\mathrm{s}} \equiv \frac{R_{\mathrm{s}}}{R_{\mathrm{c}}}-1,\,
    \label{eq:forward-shock-thickness}
\end{equation}
is nearly constant. Conversely, the reverse shock gradually separates from the contact discontinuity, so that the relative thickness of the reverse-shocked shell,
\begin{equation}
    \Delta_{\mathrm{r}} \equiv 1 - \frac{R_{\mathrm{r}}}{R_{\mathrm{c}}}\,,
\end{equation}
is fundamentally time-dependent and evolves in a way that is set primarily by the smallness parameter
\begin{equation}
    \delta(R_{\mathrm{c}}) \equiv \left( \frac{\rho_{a,0}}{\rho_{w,0}} \right)^{1/2}\left[ \frac{R_{\mathrm{c}}(t)}{R_{w,0}} \right]^{(2-n)/2}\,.
\end{equation}
In the regime of interest, $\delta \ll 1$, and the relative shell thickness $\Delta_{\rm r}$ remains small. In this regime, $\Delta_{\mathrm{r}}$ can also be expressed as
\begin{equation}
    \Delta_{\mathrm{r}} = \kappa_{\mathrm{r}}\delta(R_c)\,,
\end{equation}
where $\kappa_{\mathrm{r}}$ is a dimensionless constant set by the self-similar solution.

Despite the asymmetry between the forward- and reverse-shocked regions, the shocked fluid variables within both the forward- and reverse-shocked shells admit approximately self-similar forms. Specifically, in the forward-shocked shell,
\begin{equation}
\begin{split}
    v_{\mathrm{s}}(r,t) &= V_{\mathrm{c}}f_{\mathrm{s}}(\eta)\,,\\
    \rho_{\mathrm{s}}(r,t) &= \rho_{a,0}\left( \frac{R_{\mathrm{c}}}{R_{w,0}} \right)^{-n}g_{\mathrm{s}}(\eta)\,,\\
    p_{\mathrm{s}}(r,t) &= \rho_{a,0} V_{\mathrm{c}}^2\left( \frac{R_{\mathrm{c}}}{R_{w,0}} \right)^{-n}h_{\mathrm{s}}(\eta)\,, \label{eq:int-sssols}
\end{split}
\end{equation}
and in the reverse-shocked shell,
\begin{equation}
\begin{split}
    v_{\mathrm{r}}(r,t) &= V_{\mathrm{c}}\{1+\Delta_{\mathrm{r}}f_{\mathrm{r}}(\eta)\}\,,\\
    \rho_{\mathrm{r}}(r,t) &= \rho_{w,0}\left( \frac{R_{\mathrm{c}}}{R_{w,0}} \right)^{-2}g_{\mathrm{r}}(\eta)\,,\\
    p_{\mathrm{r}}(r,t) &= \rho_{w,0}\left( \frac{R_{\mathrm{c}}}{R_{w,0}} \right)^{-2}V_{\mathrm{c}}^2\Delta_{\mathrm{r}}^2h_{\mathrm{r}}(\eta)\,,
\end{split}
\end{equation}
Here $V_{\mathrm{c}} = dR_{\mathrm{c}}/dt$, and $\eta = (r - R_{\mathrm{c}}) / (R_{\mathrm{c}} - R_{\mathrm{r}})$ in the reverse-shocked shell, while $\eta = (r - R_{\mathrm{c}}) / (R_{\mathrm{s}} - R_{\mathrm{c}})$ in the forward-shocked shell. Substituting these expressions into the Euler equations reduces the flow to a coupled system of ordinary differential equations for the dimensionless functions $f_{\mathrm{s,r}}$, $g_{\mathrm{s,r}}$, and $h_{\mathrm{s,r}}$, which determine the structure of the shocked gas in each region. We make use of additional corrections for the forward-shocked variables to account for the effects of finite $f$, outlined in \cite{2024ApJ...975L..14C}.

In these solutions, the position of the contact is given explicitly by
\begin{equation}
    R_{\mathrm{c}}(t) = R_{w}[1-\kappa_{\mathrm{c}}\delta(R_{\mathrm{c}})]\,,
\end{equation}
where $R_w = R_{w,0} + v_w t$, and $\kappa_{\mathrm{c}}$ is a dimensionless constant fixed by the similarity solution.

\subsection{Transition to energy-conserving phase}
\label{subsec:characteristic-timescales}

The interaction solution remains valid so long as the reverse-shocked region remains thin relative to the forward-shocked shell. A natural estimate for the end of this phase is obtained by requiring that the reverse-shocked shell no longer remain thinner than the forward-shocked shell, i.e. $\Delta_{\mathrm{r}} \sim \Delta_{\mathrm{s}}$. Letting $t_{\mathrm{ss,end}}$ represent the time at which this occurs, then
\begin{equation}
    \kappa_{\mathrm{r}}\delta(t_{\mathrm{ss,end}}) = \Delta_{\mathrm{s}}\,.
\end{equation}
Using the expression for $\delta(t)$ outlined earlier, this becomes
\begin{equation}
    \frac{R_{w}(t_{\mathrm{ss,end}})}{R_{w,0}} = \left[ \frac{\Delta_{\mathrm{s}}}{\kappa_{\mathrm{r}}}\left( \frac{\rho_{w,0}}{\rho_{a,0}} \right)^{1/2} \right]^{2/(2-n)}\,.
\end{equation}
Assuming $R_{w}(t_{\mathrm{ss,end}}) \sim v_wt_{\mathrm{ss,end}}$, one finds
\begin{equation}
    t_{\mathrm{ss,end}} = \frac{R_{w,0}}{v_w}\left( \frac{\Delta_{\mathrm{s}}}{\kappa_{\mathrm{r}}} \right)^{2/(2-n)} \left( \frac{\rho_{w,0}}{\rho_{a,0}} \right)^{1/(2-n)}\,.
    \label{eq:tss}
\end{equation}
For $n < 2$, this yields a finite transition time at which the interaction solution ceases to apply. As $n \rightarrow 2$, the exponent diverges, indicating that the interaction regime no longer transitions to a distinct later energy-conserving phase.

Once the reverse-shocked shell width becomes comparable to the width of the forward-shocked shell, or equivalently, once the density of the swept ambient medium becomes comparable to that of the shocked wind, the flow approaches the pressure-driven, energy-conserving regime first \textbf{analysed} by \citet{1972SvA....15..708A, 1975A&A....43..323F, 1977ApJ...218..377W}, in which the shocked wind remains hot and approximately isobaric, and the expansion is driven primarily by the conversion of wind kinetic energy into thermal energy at a radius deeply interior to the contact. For a power-law ambient medium, the corresponding generalization of the forward-shock evolution was given by \cite{1988RvMP...60....1O}, while the reverse-shock scaling was discussed in \cite{1992ApJ...388..103K}.

The temporal scaling of the forward shock (and contact discontinuity) in this regime can be derived by assuming that 1) the rate at which energy is injected into the forward-shocked shell scales in proportion to the (assumed time-independent) energy injection rate from the wind, which should be upheld when $R_{\rm r} \ll R_{\rm c}$ (or when the sound speed of the shocked ejecta is large in comparison to the speed of the reverse shock itself; see \citealt{1977ApJ...218..377W}), 2) the pressure at the contact discontinuity scales with the post-forward-shock pressure, and 3) the radius of the contact discontinuity scales in proportion to the radius of the forward shock. Taken together, these assumptions predict
\begin{equation}
\begin{split}
    \dot{E} &= 4\pi R_{\rm c}^2V_{\rm c} p_{\rm c} \propto R_{\rm s}^2 V_{\rm s} R_{\rm s}^{-n}V_{\rm s}^2 = {\rm const.} \\
    &\Rightarrow V_{\rm s} \propto R_{\rm s}^{(n-2)/3},
\end{split}
\end{equation}
such that the temporal power-law solutions for the radius of the forward shock and the pressure of the shocked wind (equal to the pressure at the contact discontinuity) are given by
\begin{equation}
    R_{\mathrm{s}}(t) \propto t^{3/(5-n)}\,, \hspace{0.5cm} p_{\rm c}(t) \propto t^{-(4+n)/(5-n)}\,.
    \label{eq:weaver-scalings}
\end{equation}

Within the shocked ambient shell, the flow admits the usual self-similar structure, which can be defined in terms of the standard self-similar variable $\xi = r/R_{\rm s}(t)$ that treats the forward shock radius as the fundamental radial scale. On the other hand, because we know that the solution must terminate in a contact discontinuity, it is more natural to introduce the same self-similar variable that is used for the interaction phase, i.e., the energy-conserving solutions for the forward shock can be written in precisely the same way as Equations \eqref{eq:int-sssols}. Because this simply corresponds to a change of variables, the functions are the same as those derived in \citet{1977ApJ...218..377W}, now written in terms of $\eta$ and rescaled by the appropriate factors of $V_{\rm c}/V_{\rm s}$.

The structure of the shocked wind can be determined under the isobaric assumption: with the temporal scaling of the pressure given in Equation \eqref{eq:weaver-scalings}, the gas-energy equation becomes
\begin{equation}
    \frac{1}{r^2}\frac{\partial}{\partial r}\left[r^2 v_{\mathrm{r}}\right] = -\frac{1}{\gamma}\frac{1}{p_{\rm c}}\frac{\partial p_{\rm c}}{\partial t} = \frac{n+4}{3\gamma}\frac{V_{\rm c}}{R_{\rm c}},
\end{equation}
which is integrated alongside the boundary condition $v_{\mathrm{r}}(R_{\rm c}) = V_{\rm c}$ to give
\begin{equation}
    v_{\mathrm{r}} = V_{\rm c}\left(\frac{n+4}{9\gamma}\frac{r}{R_{\rm c}}+\left(1-\frac{n+4}{9\gamma}\right)\left(\frac{r}{R_{\rm c}}\right)^{-2}\right). \label{eq:vprof}
\end{equation}
This shows that the velocity structure of the shocked wind is self-similar, and depends only on $r/R_{\rm c}$. Since the reverse shock has already been assumed to be at small radii relative to the contact discontinuity, this also shows 
\begin{equation}
    v(R_{\rm r}) \simeq \left(1-\frac{n+4}{9\gamma}\right)V_{\rm c}\left(\frac{R_{\rm r}}{R_{\rm c}}\right)^{-2} \simeq \frac{\gamma-1}{\gamma+1}v_{\rm w},
\end{equation}
where the final equality follows from the jump conditions at the reverse shock and the assumption that the reverse shock speed is much less than the wind speed. The radius of the reverse shock therefore satisfies
\begin{equation}
    R_{\rm r} = R_{\rm c}\sqrt{\frac{\gamma+1}{\gamma-1}\left(1-\frac{n+4}{9\gamma}\right)}\sqrt{\frac{V_{\rm c}}{v_{\rm w}}} \propto R_{\rm c}^{\frac{n+4}{6}} \propto t^{\frac{n+4}{2\left(5-n\right)}}, \label{eq:Rr}
\end{equation}
where the last proportionality follows from Equation \eqref{eq:weaver-scalings} (recall that the radius of the contact discontinuity is assumed to scale in proportion with the radius of the forward shock). 

Finally, we can determine the density profile of the shocked wind by noting that the boundary condition at the reverse shock is $\rho(R_{\rm r}) \propto R_{\rm r}^{-2} \propto R_{\rm c}^{-(n+4)/3}$. Since the velocity profile \eqref{eq:vprof} is self-similar and $R_{\rm r}/R_{\rm c} \simeq 0$ in this regime, the density permits a self-similar solution of the form
\begin{equation}
    \rho_{\mathrm{r}} = \frac{\gamma+1}{\gamma-1}\rho_{w,\mathrm{r}}\left(\frac{R_{\rm c}}{R_{\rm c, 0}}\right)^{-\frac{n+4}{3}}g\left(\frac{r}{R_{\rm c}}\right),
\end{equation}
where $R_{\rm c, 0}$ is the radius of the contact discontinuity at some initial time, $\rho_{w,\mathrm{r}}$ is the wind density at the reverse shock when $R_{\rm c}$ coincides with that scale radius, and $g(0) = 1$ to satisfy the boundary condition at the reverse shock (again assuming $R_{\rm r}/R_{\rm c} \simeq 0$). Then using the velocity profile in the continuity equation shows that 
\begin{equation}
    g\left(\frac{r}{R_{\rm c}}\right) = \left(1-\frac{r^3}{R_{\rm c}^3}\right)^{-\frac{\left(n+4\right)\left(\gamma-1\right)}{9\gamma-n-4}},
\end{equation}
and hence the density is
\begin{equation}
\begin{split}
    \rho_{\mathrm{r}} &= \frac{\gamma+1}{\gamma-1}\rho_{w,\mathrm{r}}\left(\frac{R_{\rm c}(t)}{R_{\rm c, 0}}\right)^{-\frac{n+4}{3}}\left(1-\frac{r^3}{R_{\rm c}(t)^3}\right)^{-\frac{\left(n+4\right)\left(\gamma-1\right)} {9\gamma-n-4}} \\
    &\propto t^{-\frac{n+4}{5-n}}\left(1-\frac{r^3}{R_{\rm c}(t)^3}\right)^{-\frac{\left(n+4\right)\left(\gamma-1\right)} {9\gamma-n-4}}.
\end{split}
\end{equation}

These solutions were presented in \cite{1977ApJ...218..377W} for the specific case of $n=0$ and $\gamma=5/3$; while the velocity profile was later generalized in \cite{1992ApJ...388..103K}, we have not encountered these generalizations for $\rho$ elsewhere. Note that they break down for $n \ge 2$, as they are based on the assumption that $R_{\rm r} \ll R_{\rm c}$, which -- from Equation \eqref{eq:Rr} -- will only be upheld when $n < 2$. They also illustrate that, in the limit that the radius of the reverse shock is much less than the contact discontinuity, the inner shocked-wind flow is \emph{also self-similar} in terms of the same variable as the interaction phase for $R_{\rm r} < r <R_{\rm c}$. Specifically, because $R_{\rm r} = 0$ is just a limiting case, the preceding solutions can be written
\begin{equation}
        p_{\mathrm{r}} = \rho_{a,0} V_{\rm c}^2\left(\frac{R_{\rm c}}{R_{\rm c, 0}}\right)^{-n}h_{\rm s}(0),
\end{equation}
\begin{multline}
        v_{\mathrm{r}} = V_{\rm c}\bigg(\frac{n+4}{9\gamma}\left(1+\left(1-\frac{R_{\rm r}}{R_{\rm c}}\right)\eta\right) \\ +\left(1-\frac{n+4}{9\gamma}\right)\left(1+\left(1-\frac{R_{\rm r}}{R_{\rm c}}\right)^{-2}\right)\bigg)
\end{multline}
\begin{multline}
        \rho_{\mathrm{r}} = \frac{\gamma+1}{\gamma-1}\rho_{w,\mathrm{r}}\left(\frac{R_{\rm c}}{R_{\rm c, 0}}\right)^{-\frac{n+4}{3}} \\ 
        \times\left(1-\left(1+\left(1-\frac{R_{\rm r}}{R_{\rm c}}\right)\eta\right)^3\right)^{-\frac{\left(n+4\right)\left(\gamma-1\right)}{9\gamma-n-4}},
\end{multline}
where $h_{\rm s}(0)$ is the self-similar pressure at the contact discontinuity as determined from the forward shock solution, and $\eta$ is the same variable as introduced in Section~\ref{subsec:interaction-phase} for the reverse-shock shell. This demonstrates that the Weaver solution is simply the one that approximates $R_{\rm r} = 0$, and this representation can be used to directly compare to the interaction-phase solutions given any (finite) value of $R_{r}/R_{c}$.

As an aside, the equivalent parametrization (i.e., substitute the standard self-similar variable $\xi$ for $\eta$) can be used for the Sedov-Taylor blastwave \citep{1946JApMM..10..241S,1950RSPSA.201..159T} in scenarios where the contact discontinuity approaches the origin (relatively small values of $n$ for a given $\gamma$; see \citealt{1990ApJ...358..214G}). Analogously to the present case where the reverse shock reaching the origin constitutes the ``true'' self-similar solution that is reached at asymptotically late times, the exact Sedov-Taylor solution holds when the contact discontinuity reaches the origin. Nevertheless, and just as the self-similar velocity profile of the shocked wind can be used to constrain the propagation rate of the reverse shock, i.e., Equation \eqref{eq:Rr}, the velocity profile of the Sedov-Taylor blastwave can be used to predict the propagation rate of the contact discontinuity: in the limit that $r\rightarrow 0$, the Sedov-Taylor velocity profile satisfies 
\begin{equation}
    v(r,t) = \frac{2}{\gamma+1}V_{\rm s}(t)\frac{r}{R_{\rm s}(t)}.
\end{equation}
Because the fluid velocity must be continuous across the contact discontinuity, it follows that its velocity and time dependent position are
\begin{equation}
    V_{\rm c} = \frac{2}{\gamma+1}\frac{V_{\rm s}}{R_{\rm s}}R_{\rm c} \,\,\, \Rightarrow \,\,\, R_{\rm c} \propto R_{\rm s}^{\frac{2}{\gamma+1}} \propto t^{\frac{4}{\left(5-n\right)\left(\gamma+1\right)}}.
\end{equation}
Thus, and even though this is a distinct power-law dependence from the forward shock, the time-dependent evolution of the contact discontinuity can be gleaned from the approximate self-similar behaviour of the shocked ambient fluid.

\section{Hydrodynamic simulations}

\subsection{Numerical methodology}
\label{sec:numerical-methods}

Although this work concerns the properties of non-relativistic winds ($u_w \lesssim 0.01$), our numerical simulations employ a one dimensional finite-volume special relativistic hydrodynamics (SRHD) code originally developed for modelling gamma-ray burst afterglows, as its moving-mesh framework proves particularly well-suited to this problem. Specifically, the domain boundaries are designed to track discontinuities (shocks and contact discontinuities), which ensures consistently high resolution in the narrow shocked regions without wasting grid cells on the unshocked wind and ambient medium, both of which are analytically prescribed. The details of this code, including the moving mesh algorithm, are presented in Appendix~\ref{sec:srhd_code}.

All simulations involve the collision of a cold, spherical, constant-velocity wind ($u_w=0.01$) with a cold, power-law ambient medium, initialized at radius $R_{w,0}=1$ and time $t_{w,0}=1$. It is useful to express $t$ in dynamical times $t_{\mathrm{dyn}}=R_{w,0}/u_w$, noting that for our parameter choices, $t_{\mathrm{dyn}}=100\,t_{w,0}$. The ambient medium and unshocked wind densities are defined according to Section~\ref{sec:basic-considerations}.

The discontinuities that bound the subdomains are initialized by solving a two-shock Riemann problem \citep{2005MNRAS.363...93Z} over a small interval of time $\delta t$, where $\delta t / t_{w,0} = 10^{-6}$. The domain edges then consist of the reverse shock at $R_{\mathrm{r}}$, the contact discontinuity at $R_{\mathrm{c}}$, and the forward shock at $R_{\mathrm{s}}$, with $R_{\mathrm{r}} < r < R_{\mathrm{c}}$ comprising the reverse-shocked subdomain, and $R_{\mathrm{c}} < r < R_{\mathrm{s}}$ comprising the forward-shocked subdomain. At the inner boundary, inflow conditions are imposed using the wind primitive state $(\rho_w, u_w, p_w)$ evaluated at $R_{\mathrm{r}}$. At the outer boundary, the ambient medium state $(\rho_a, u_a, p_a)$ is prescribed at $R_{\mathrm{s}}$.

We performed 25 simulations with $\gamma=5/3$, $n \in [0, 0.5, 1, 1.5, 2]$ and $f \in [10^1, 10^2, 10^3, 10^4, 10^5]$. The two shocked subdomains were resolved with 800 zones each, which was sufficient to resolve the overall profiles; see Appendix~\ref{sec:numerical-convergence} for a discussion of convergence. Each simulation was evolved to $t/t_{\mathrm{dyn}}=10^{18}$.

\subsection{Results}
\label{sec:results}

\subsubsection{Interaction phase}
\label{subsec:interaction-regime}

\begin{figure*}
\begin{center}
\includegraphics{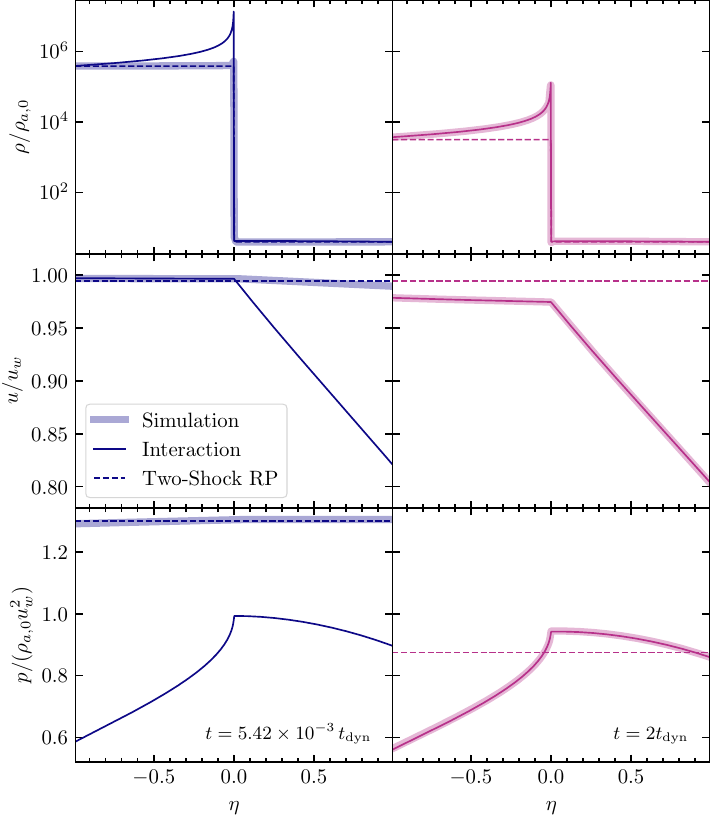}
\caption{Comparisons of our simulated fluid profiles for $f=10^5$ to those of the interaction solutions and of the two-shock Riemann problem (RP) solutions at two dynamical times (\textbf{right}) and at $t \ll t_{\mathrm{dyn}}$ (\textbf{left}). For $t \ll t_{\mathrm{dyn}}$, only minimal radial structure has formed in each of the shocked shells; as such, the system is well-approximated by the solution to the two-shock RP. By $t\sim t_{\mathrm{dyn}}$, gradients have steepened in every fluid variable within each shocked shell, and the simulation has fully relaxed into the interaction regime.}
\label{fig:two_zone}
\end{center}
\end{figure*}

In the limit $f \gg 1$, the early-time evolution should be described by the interaction phase solutions presented in \cite{2024ApJ...975L..14C}, following a transient that originates from the initial conditions -- i.e., the state first has to ``relax'' from the two-shock Riemann problem (RP) to the self-similar solution. Given the initial rate at which the CD-RS shell expands, the estimate for when the self-similar state is reached is $t_{\mathrm{dyn}}$ (see the discussion in Section 4.1 of \citealt{2024ApJ...975L..14C}). Figure~\ref{fig:two_zone} shows convergence to the interaction solutions as the system evolves from $t \ll t_{\mathrm{dyn}}$ to $t \sim t_{\mathrm{dyn}}$, also depicting the simultaneous divergence from the two-shock RP solution. This transition from the initial two-shock RP to the interaction regime occurs many orders of magnitude earlier than the transition from the interaction to energy-conserving stage, which occurs approximately when the wind density declines to that of the ambient density (see Section~\ref{subsec:characteristic-timescales}).

\begin{figure}
\begin{center}
\includegraphics{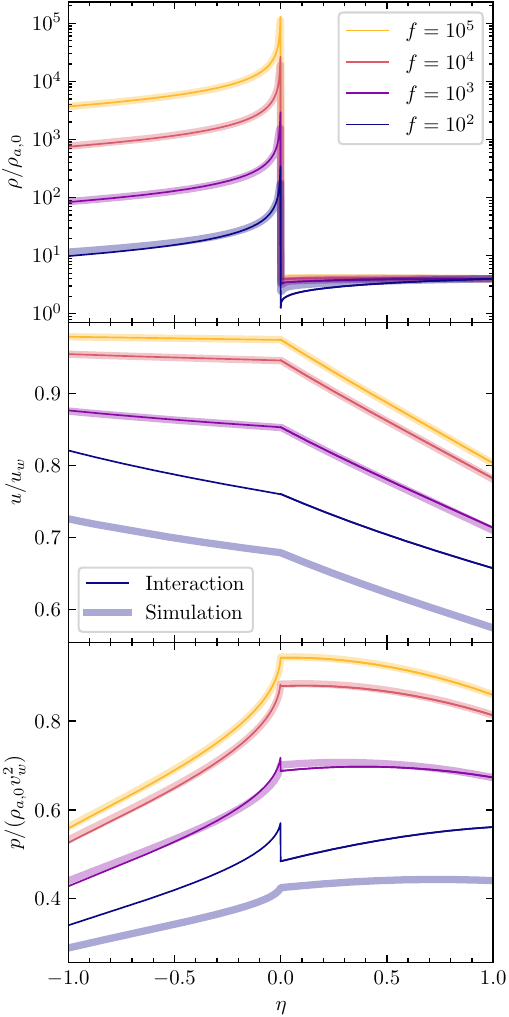}
\caption{Comparison of the density (\textbf{top}), velocity (\textbf{middle}), and pressure (\textbf{bottom}) profiles from our simulations to those predicted by the interaction solutions for $n=0$. The representative time at which each snapshot was taken for any given $n$ corresponds to the minimum relative $L_1$ error of the forward-shocked shell pressure between the simulation output and the interaction-phase self-similar solutions. Since these solutions assume $f \gg 1$, results converge as $f$ increases.}
\label{fig:f-convergence}
\end{center}
\end{figure}

Figure~\ref{fig:f-convergence} shows the convergence of our simulations and the interaction solutions for different $f$ at $t\sim t_{\mathrm{dyn}}$ in the case of $n=0$: as the overdensity parameter is increased from $f=10^2$ to $10^5$, the simulated profiles approach the interaction solutions throughout the shocked structure. For all other diagrams in this work, we use $f=10^5$ to ensure sufficient convergence to the self-similar solutions in the interaction regime.

\begin{figure}
\begin{center}
\includegraphics{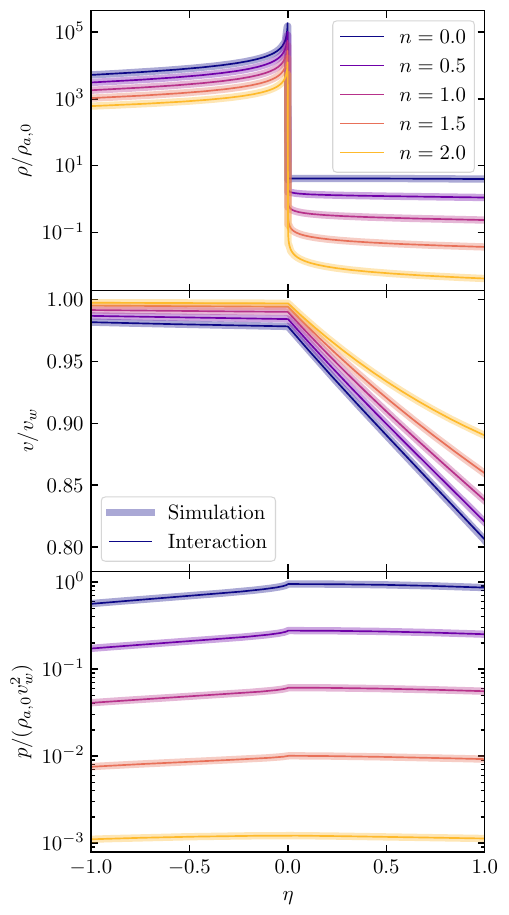}
\caption{Radial fluid profiles from our simulations and the interaction solutions for $f=10^5$ and $n\in [0, \frac{1}{2}, 1, \frac{3}{2}, 2]$. The representative time at which each snapshot was taken for any given $n$ corresponds to the minimum relative $L_1$ error of the forward-shocked shell velocity between the simulation output and the interaction-phase self-similar solutions. Agreement is robust across all tested ambient medium density profiles.}
\label{fig:interaction-phase}
\end{center}
\end{figure}

This agreement extends to all other $n$ tested, as shown in Figure~\ref{fig:interaction-phase}. For each value of $n$, the snapshot across all three fluid variables ($\rho$, $v$, and $p$) was chosen at the time for which the relative $L_1$ error in the forward-shocked velocity profile, measured with respect to the interaction self-similar solution, is minimized. Across the full range of ambient density profiles considered here, the interaction solutions effectively capture the structure of the simulated profiles in both shocked regions given $f \gg 1$.

\subsubsection{Transition to energy-conserving phase}
\label{subsec:transition-interaction}

The shocked regions only remain in the interaction regime so long as the reverse-shocked shell remains thin compared to the forward-shocked shell. Once these shell widths become comparable, so too do the wind and ambient densities, and the flow is expected to depart from this interaction-dominated state and resemble the later, energy-conserving regime. To quantify this transition in the simulations, we compare the radial pressure profile in the forward-shocked shell to both sets of self-similar solutions and compute the relative $L_1$ error as a function of time.

\begin{figure*}
\begin{center}
\includegraphics{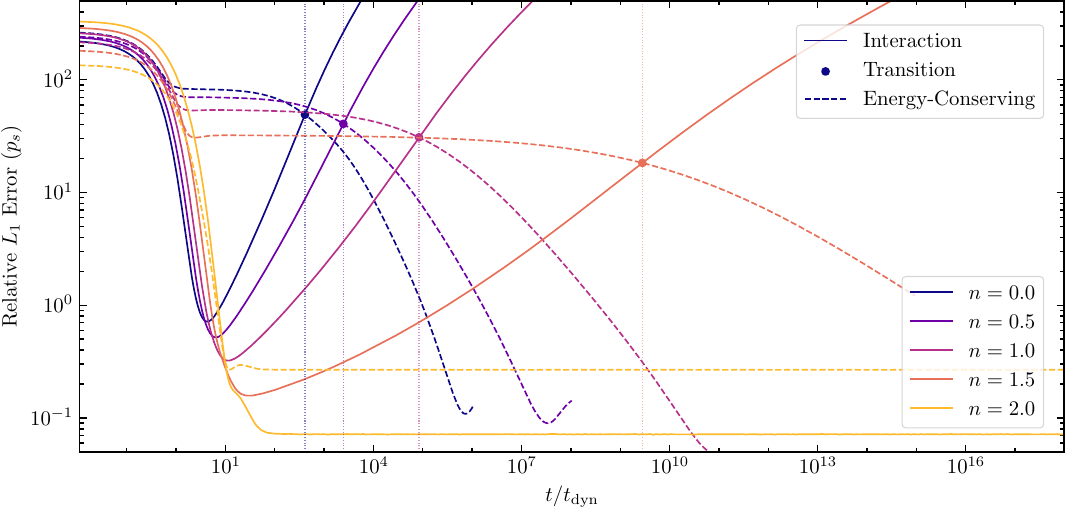}\\[2.5em]
\includegraphics{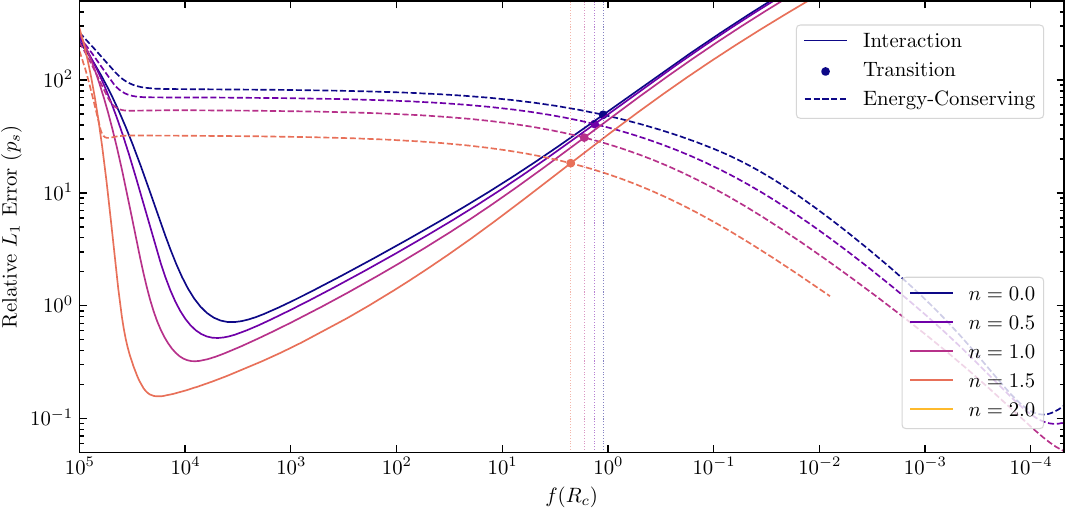}
\caption{Relative $L_1$ error between our hydrodynamics simulations and self-similar solutions in both interaction and energy-conserving regimes for the radial pressure profiles in the forward-shocked shell. The initial overdensity parameter $f$ was set to $10^5$ for each case, and $n \in [0,\frac{1}{2},1,\frac{3}{2},2]$. The change in error is shown vs. time (\textbf{top}) and instantaneous overdensity $f(R_{\mathrm{c}})\equiv f\cdot (R_{w,0}/R_{\mathrm{c}})^{2-n}$ (\textbf{bottom}). We measure $t_{\mathrm{ss,end}}$ to be the time at which these error curves intersect. This transition from the interaction to energy-conserving stage takes place at various times for different $n$, but at comparable $f(R_{\mathrm{c}}) \sim 1$. For $n=2$, no such transition occurs since the system maintains a perpetual similarity state with $f$ constant throughout its evolution.}
\label{fig:l1-error}
\end{center}
\end{figure*}

The top panel of Figure~\ref{fig:l1-error} shows the resulting errors vs. time for $n \in [0, \frac{1}{2}, 2, \frac{3}{2}, 2]$. At the earliest times, this error is high since we do not initialize the simulations in the interaction self-similar state; as such, the system relaxes into the interaction regime over the course of $\sim$ a few dynamical times. Once this relaxation has occurred, the interaction solutions provide highly accurate descriptions of the fluid behaviour, until the error again begins to rise as the system gradually departs from these solutions and moves toward the energy-conserving ones. We define $t'_{\mathrm{ss,end}}$ -- the measured analogue to the derived $t_{\mathrm{ss,end}}$ in Section~\ref{subsec:interaction-phase} -- to be the time at which the two error curves intersect, i.e., the epoch at which the interaction and energy-conserving descriptions are equally (in)accurate. For $n=2$, no such intersection occurs: the interaction solution remains the more accurate description throughout the evolution, consistent with the expectation that this case does not transition to a distinct later energy-conserving similarity state (see the discussion toward the end of Section~\ref{sec:basic-considerations}). For all $n$, the error does not reach 0 for two reasons: (1) the numerical simulations are not infinitely accurate, exhibiting (e.g.) discrepancies at the CD due to the finite nature of the numerical method (see Appendix~\ref{sec:numerical-convergence}), and (2) the interaction solutions used here do not include any of the higher-order terms discussed in \cite{2024ApJ...975L..14C}, i.e., the self-similar solutions are only the leading-order terms in an infinite series in $\delta$.

The bottom panel of Figure~\ref{fig:l1-error} shows the L1 error vs.~instantaneous $f(R_c) = f \cdot (R_{w,0}/R_c)^{2-n}$ for $n \in [0, \frac{1}{2}, 2, \frac{3}{2}, 2]$. The transition from the interaction to energy-conserving stage occurs around $f(R_c) \sim 1$ for all $n < 2$, demonstrating that this change takes place when the densities of the wind and ambient medium are comparable. This is an expected result since $f(R_\mathrm{c}) = 1/\delta^2(R_{\mathrm{c}}) = (\kappa_{\mathrm{r}} / \Delta_{\mathrm{r}})^2$, and because $\Delta_{\mathrm{s}} \sim \Delta_{\mathrm{r}}$ at the end of the interaction regime, $f(R_\mathrm{c}) \sim (\kappa_{\mathrm{r}} / \Delta_{\mathrm{s}})^2 \sim 1$.

\begin{figure}
\begin{center}
\includegraphics{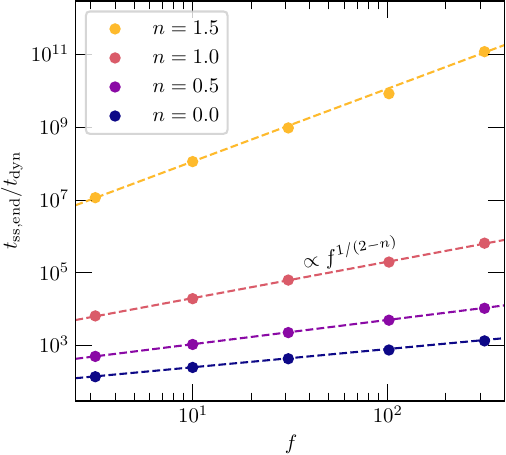}
\caption{The duration $t'_{\mathrm{ss,end}}$ of the interaction phase as a function of initial overdensity parameter $f$. The duration is proportional to $f^{1/(2-n)}$ as predicted by Equation~\ref{eq:tss}, and is therefore extended for steeper ambient medium density profiles. For $n=2$, the system is characterized by a single perpetual self-similar state, corresponding to $t_{\mathrm{ss,end}} \rightarrow \infty$.}
\label{fig:timescales}
\end{center}
\end{figure}

The measured dependence of $t'_{\mathrm{ss,end}}$ on the overdensity parameter is shown in Figure~\ref{fig:timescales}. For each $n < 2$, the measured transition time increases with both $f$ and $n$, following the scaling $t_{\mathrm{ss,end}} \propto f^{1/(2-n)}$ predicted by Equation~\ref{eq:tss}. The duration of the interaction phase therefore grows rapidly with increasing ambient density slope, and as $n \rightarrow 2$, the exponent in the predicted scaling relation diverges, while the measured behaviour approaches a single persistent similarity state. Thus, the interaction regime can remain dynamically important for a substantial fraction of the pre-radiative evolution, especially in steep ambient density profiles.

Furthermore, depending on the criterion for an acceptable threshold of error, Figure~\ref{fig:l1-error} shows that for $n < 2$, the window of time in which the system is not well-modelled by either set of self-similar solutions is relatively broad. Even adopting a conservative criterion in which a model is regarded as ``valid'' once its relative error lies within an order of magnitude of its global minimum, the system can remain in this intermediate phase until times several orders of magnitude larger than the characteristic transition timescale itself. This indicates that the transition between the interaction and energy-conserving regimes is itself dynamically extended, and it may not be a suitable approximation to model the system with an abrupt change from one asymptotic solution to the other, though extending the interaction solutions to higher-order terms may shorten this transition window (as discussed further in Section~\ref{sec:conclusions}). \textbf{Additionally, we interpolate between these two regimes and provide analytic time-dependent functions for the forward shock, contact discontinuity, and reverse shock positions in Appendix~\ref{sec:analytic-functions}}.

\begin{figure*}
\begin{center}
\includegraphics{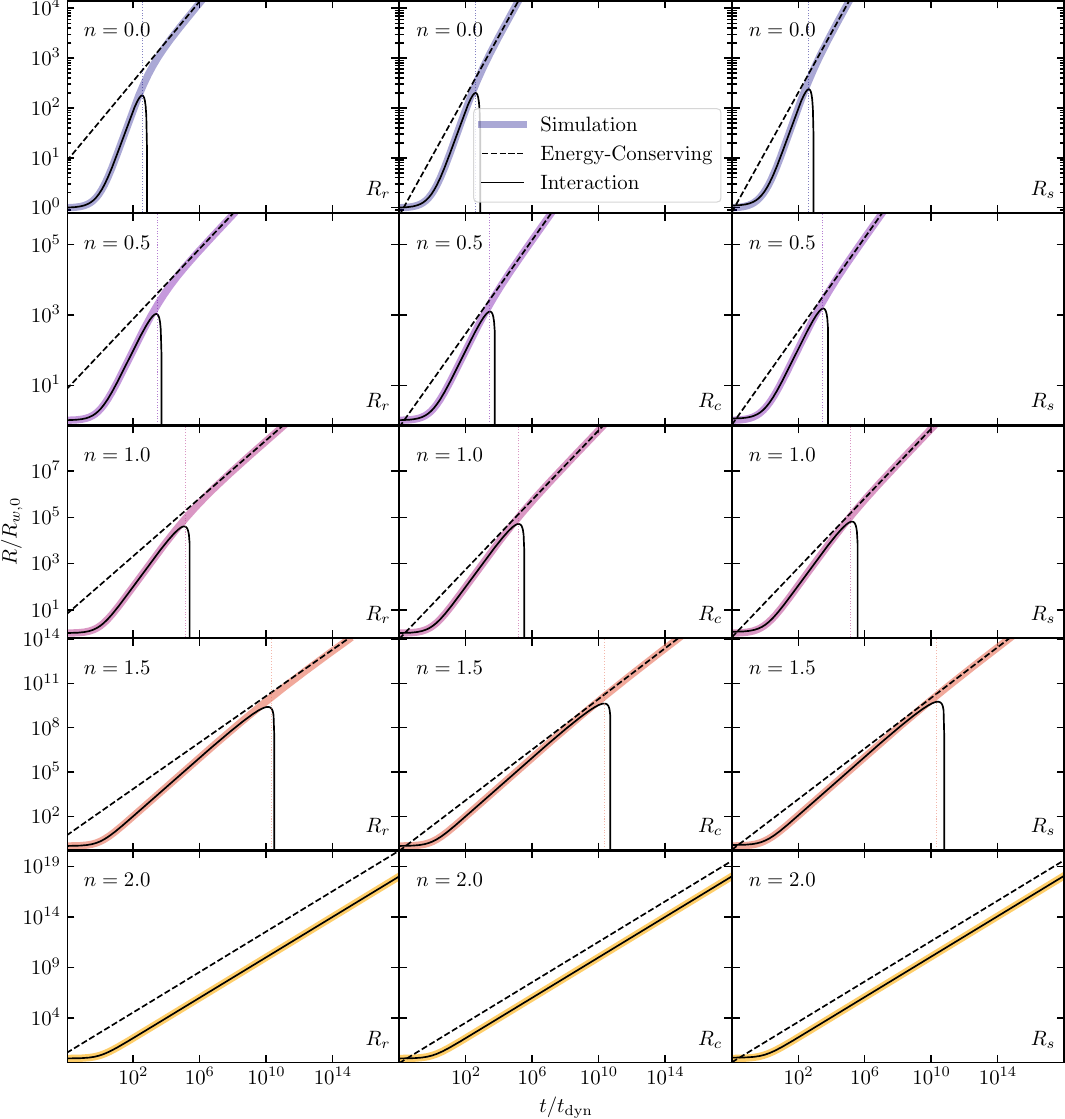}
\caption{The positions of the reverse shock, contact discontinuity, and forward shock as a function of time for different values of $n$ from our hydrodynamic simulations, the interaction solutions, and the energy-conserving solutions. The dotted vertical lines represent the measured timescales characterizing the transitions from the interaction to energy-conserving regime. For $n=2$, these are not distinct regimes, and the system exists in a single perpetual similarity state.}
\label{fig:contact-comparisons}
\end{center}
\end{figure*}

As the shocked regions evolve away from the interaction phase and into the energy-conserving phase, the fluid profiles and shock structure change accordingly. Figure~\ref{fig:contact-comparisons} shows the trajectories of the reverse shock, contact discontinuity, and forward shock for each value of $n$, together with the corresponding interaction and energy-conserving solutions. At early times, all three discontinuities are well described by the interaction-stage solutions. For $n < 2$, the simulated shock structure subsequently departs from this early-time behaviour at $t \sim t_{\mathrm{ss,end}}$ and approaches the later energy-conserving scalings, adhering specifically to the power-law shock temporal scalings in Equations~\ref{eq:weaver-scalings} \& \ref{eq:Rr}. For $n=2$, the discontinuity positions remain consistent with a single similarity state, and differ in normalization from the energy-conserving predictions since the system never leaves the interaction regime.

\begin{figure*}
\begin{center}
\includegraphics{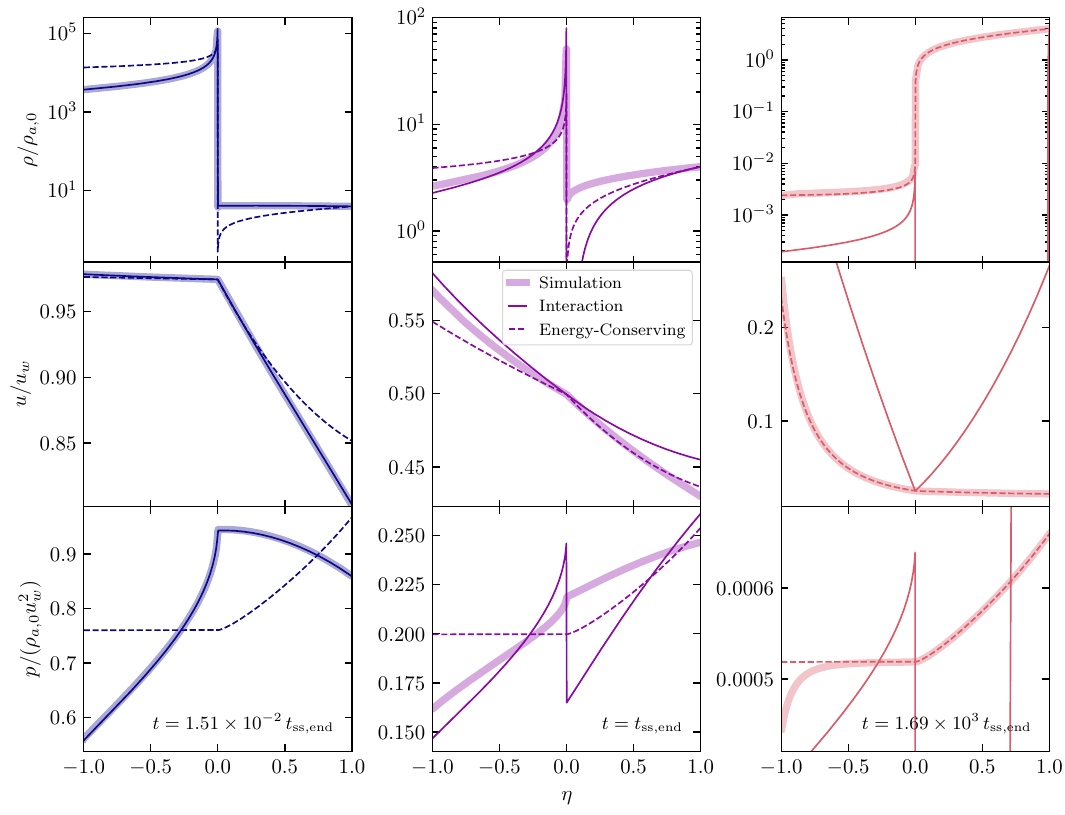}
\caption{Comparisons of our simulated fluid profiles to those of the interaction and energy-conserving solutions across various epochs in the system's evolution. The initial overdensity ratio $f=10^5$ for this example, and the ambient medium density profile is uniform ($n=0$). The snapshots here correspond to instantaneous overdensities of $f(R_{\mathrm{c}})=9.227\times 10^2$ (\textbf{left}), $f(R_{\mathrm{c}})=1.109$ (\textbf{middle}), and $f(R_{\mathrm{c}})=4.843\times 10^{-5}$ (\textbf{right}), again reinforcing the criterion that the transition period is characterized by $f(R_{\mathrm{c}}) \sim 1$. During the interaction ($t \ll t_{\mathrm{ss,end}}$) and energy-conserving ($t \gg t_{\mathrm{ss,end}}$) phases, the radial profiles of all fluid variables are well-modelled by one set of self-similar solutions, while the transition phase at $t \sim t_{\mathrm{ss,end}}$ is not well-modelled by either set.}
\label{fig:snapshots}
\end{center}
\end{figure*}

The evolution of the fluid profiles is shown explicitly in Figure~\ref{fig:snapshots} for the representative case $n=0$, $f=10^5$, at epochs before, during, and after the transition. At early times ($t \ll t_{\mathrm{ss,end}}$), the simulated profiles are well described by the interaction solution across both shocked regions, and at late times ($t \gg t_{\mathrm{ss,end}}$), the same quantities are instead well reproduced by the energy-conserving solution. Near $t \sim t_{\mathrm{ss,end}}$, however, neither asymptotic description provides a comparably good match to the full shocked structure.

\section{Discussion}
\label{sec:discussion}

\subsection{Duration of the energy-conserving phase}
\label{subsec:duration-energy-conserving}

The energy-conserving phase lasts as long as radiative losses remain dynamically negligible -- e.g. as long as the forward-shocked ambient shell cannot cool on a timescale comparable to the expansion time. Beyond this point, the shell rapidly compresses into a thin, dense layer, while the reverse-shocked wind interior remains hot, marking the onset of the first radiative stage described in \cite{1977ApJ...218..377W}.

An equivalent way to estimate this transition is through the comparison of the rate at which the contact discontinuity does work on the swept-up shell (i.e., the rate at which energy is added to the shocked ambient gas via the contact discontinuity) to the rate at which the shell radiates energy. The former rate is, from energy conservation and the isobaric structure of the shocked wind,
\begin{equation}
    \dot{E}_{\mathrm{c}} = \frac{9(\gamma-1)}{9\gamma-n-4}L_w\,,
\end{equation}
while the radiative luminosity $L_{\mathrm{rad}}$ of the forward-shocked shell may be estimated as
\begin{equation}
    L_{\mathrm{rad}} \sim n_{\mathrm{s}}^2\Lambda(T_{\mathrm{s}})\mathcal{V}_{\mathrm{s}}\,,
\end{equation}
where $n_{\mathrm{s}}$ and $T_{\mathrm{s}}$ are the characteristic post-shock number density and temperature, $\mathcal{V}_{\mathrm{s}}$ is the shell volume, and $\Lambda(T_{\mathrm{s}})$ is the cooling function evaluated at the post-shock temperature. Here we take the ion, electron, and total number densities to be equal to within factors of order unity. Using the strong shock compression ratio $n_2/n_1 = (\gamma+1)/(\gamma-1)$, together with $n_{\mathrm{s}} \sim \rho_{\mathrm{s}}/m_p$, gives
\begin{equation}
    L_{\mathrm{rad}} \sim \frac{4}{3}\pi R_{w,0}^3\left(1-\xi_{\rm c}^3\right)n_{a,0}^2\Lambda(T_{\mathrm{s}}) \left( \frac{\gamma+1}{\gamma-1} \right)^2 \left( \frac{R_{\mathrm{s}}}{R_{w,0}} \right)^{3-2n}\,,
\end{equation}
where $\xi_{\rm c} = R_{\rm c}/R_{\rm s}$ is the ratio of the contact discontinuity radius to the forward shock radius, which is constant during the adiabatic (Weaver/energy-conserving) phase. The transition to the radiative stage is then estimated by requiring $L_{\mathrm{rad}} \sim \dot{E}_c$, which yields a shock position of
\begin{equation}
    \frac{R_{\mathrm{s}}}{R_{w,0}} \sim \left[ \frac{27(\gamma-1)^3L_w}{4\pi n_{a,0}^2R_{w,0}^3(1 - \xi_{\mathrm{c}}^3)\Lambda(T_{\mathrm{s}})(9\gamma-n-4)(\gamma+1)^2} \right]^{1/(3-2n)}\,.
    \label{eq:radiative-transition}
\end{equation}
This expression gives the characteristic shock radius at which the energy-conserving phase ceases to apply. For typical cooling curves, the dominant dependence enters through the drop in post-shock density with increasing radius, so the transition radius scales approximately as $\mathrm{const.}^{1/(3-2n)}$, with $\mathrm{const.}>1$. The transition is therefore pushed to larger radii for steeper ambient density profiles. Since the interaction phase is also lengthened for steeper ambient density profiles, the cumulative effect is that steeper ambient density profiles result in a longer total non-radiative stage, delaying the onset of the second and third stages outlined in \cite{1977ApJ...218..377W}.

Furthermore, a critical slope occurs at
\begin{equation}
    n_{\mathrm{crit}} = \frac{3}{2}\,,
\end{equation}
above which the flow is not expected to ever become dynamically dominated by radiative losses. This is particularly relevant for winds embedded in Bondi-like accretion flows, as such systems are still expected to transition to the energy-conserving regime since $n < 2$, while steeper systems where $n \geq 2$ are expected to remain interaction-dominated in perpetuity.

\subsection{Observational implications}
\label{subsec:observational-implications}

For a sufficiently overdense outflow embedded in an ambient medium with a relatively steep density gradient, our results demonstrate that the interaction phase can occupy a substantial fraction of the total non-radiative lifetime of the outflow, and in some cases may be the hydrodynamic regime most relevant to the observed emission. A useful estimate for the duration of this regime is given by Equation~\eqref{eq:tss}; since $\Delta_{\mathrm{s}} / \kappa_{\mathrm{r}}$ is of order unity, this equation can be re-expressed as
\begin{equation}
    t_{\mathrm{ss,end}} \sim \left[ \frac{v_w^{(n-3)}\dot{M}}{4\pi\rho_{a,0} R_{w,0}^n} \right]^{1/(2-n)}\,,
\end{equation}
where $R_{w,0}$ again represents the characteristic radius at which the freely expanding wind becomes established and begins driving the double-shock structure. \textbf{This transition occurs for $n < 2$ if the wind remains active indefinitely.}

\textbf{On the other hand, realistic winds are launched over a finite time interval that depends on the driving mechanism, resulting in ejecta shells of finite thickness. The dynamics of the shocked regions will then deviate from this framework once the reverse shock reaches the interior edge of the shell, which occurs at a time}
\begin{equation}
    t_{\Delta} \sim (t_{\mathrm{dyn}} + t_{\rm w,e}) f^{1/(4-n)}\,,
\end{equation}
\textbf{where $t_{\rm w,e}$ is the duration of the wind. This could be particularly important in scenarios where $t_{\rm w,e}$ is much shorter than relevant observing windows.}

\textbf{In the following subsections, we estimate the duration $t_{\mathrm{ss,end}}$ of the interaction phase for various astrophysical phenomena. For TDEs and LFBOTs specifically, depending on how long the wind is sustained, the reverse-shock crossing time may be shorter than $t_{\mathrm{ss,end}}$, though perhaps still long enough to be observationally relevant.}

\subsubsection{Massive-star winds}

For an O-type stellar wind, \cite{1975A&A....43..323F} and \cite{1977ApJ...218..377W} estimated the duration of the \textit{entire} non-radiative phase to be $\sim 2\times 10^3\,$yr, and we obtain a comparable number using similar parameters $v_w=2000\,\mathrm{km}\,\mathrm{s}^{-1}$, $\dot{M}=10^{-6}\,\mathrm{M}_{\odot}\,\mathrm{yr}^{-1}$, and $n_{a,0}=1\,\mathrm{cm}^{-3}$, and $\Lambda = 8\times 10^{-22}$ in Equation~\ref{eq:radiative-transition}, yielding $t_{\mathrm{rad}}\sim 2112\,$yr. Adopting a `wind-launch' radius of $R_{w,0}\sim 10^{12}\,\mathrm{cm}$, consistent with standard calibrations for Galactic O stars \citep{2005A&A...436.1049M}, we estimate an inner stellar wind density of $\rho_{w,0} \sim 10^{-14}\,\mathrm{g}\,\mathrm{cm}^{-3}$, which corresponds to $f \sim 10^{10}$. Since $f \gg 1$, the early wind evolution should be interaction-dominated, and we can estimate the duration of the interaction-specific component of this regime using the same parameters,
\begin{multline}
    t_{\mathrm{ss,end,O}} \sim 19.4\,\mathrm{yr} \, \times \, \left( \frac{v_w}{{2000\,\mathrm{km}\,\mathrm{s^{-1}}}} \right)^{-3/2}\times \\ \left( \frac{\dot{M}}{10^{-6}\,\mathrm{M}_{\odot}\,\mathrm{yr}^{-1}} \right)^{1/2}\times\left( \frac{n_{a,0}}{1\,\mathrm{cm}^{-3}} \right)^{-1/2}\,.
\end{multline}
Taking $R_{\mathrm{ss,end}} \equiv v_wt_{\mathrm{ss,end}}$, this corresponds to $R_{\mathrm{ss,end}}=1.73\times 10^{17}\,$cm, equivalent to the transition radius predicted in \cite{1975A&A....43..323F}, which was obtained by solving for the radius at which the unshocked wind density is equivalent to that of the unshocked ambient medium. This is consistent with our numerical result that the transition occurs when the instantaneous overdensity has declined to only $f(R_{\mathrm{c}}) \sim$ a few (Figure~\ref{fig:l1-error}), i.e. when the wind is no longer overwhelmingly denser than the ambient medium.

\subsubsection{White dwarf winds}

Another application of this framework is the white-dwarf-wind-driven remnant Pa 30 \citep{2019Natur.569..684G, 2020A&A...644L...8O}, the likely remnant of SN 1181 \citep{2021ApJ...918L..33R}. The central white dwarf appears to drive a very fast wind, $v_w \geq 10^4\,\mathrm{km}\,\mathrm{s}^{-1}$, and \cite{2026ApJ...996L...3C} argue that this wind interaction is central to the remnant's unusual radial filamentary morphology. Although we only consider spherically symmetric winds here, and therefore cannot describe the Rayleigh-Taylor growth invoked in that work, their interpretation is nevertheless embedded in the same interaction-phase framework considered here. Filament growth shuts off once the wind and ambient densities become comparable near the contact discontinuity, marking the end of the strongly overdense interaction regime. \cite{2026ApJ...996L...3C} use this same framework to infer a characteristic duration of $\sim 1-10\,\mathrm{yr}$ for the filament-growth phase in Pa 30, placing that system naturally within the broader class of interaction-dominated outflows discussed here.

\subsubsection{Galactic winds}

As highlighted in earlier sections, these interaction timescales increase for steeper $n$. Large-scale galactic outflows interact with the circumgalactic medium (CGM), adhering approximately to a power-law density profile of index $n=1-2$ \citep{2017ApJS..233...20L, 2018ApJ...862....3B, 2018MNRAS.478.2909S}. For such an outflow, we take $R_{w,0}$ to be the gas disk scale height $\sim 100\,\mathrm{pc}$, i.e. the characteristic radius at which clustered supernova-driven bubbles break out of the disc and begin venting into the halo as a galactic wind \citep[e.g.][]{1985Natur.317...44C, 2018MNRAS.481.3325F}. Adopting a wind velocity $v_w = 100\,\mathrm{km}\,\mathrm{s}^{-1}$ and mass flux $\dot{M}=10^{-1}\,\mathrm{M}_{\odot}\,\mathrm{yr}^{-1}$ \citep[e.g.][]{2019ApJ...884...53F}, as well as a local density normalization $n_{a,0} = 10^{-2}\,\mathrm{cm}^{-3}$, we estimate an initial wind overdensity $f \sim$ tens. This is a somewhat modest overdensity, implying a relatively short-lived interaction phase—however, the steepness of the ambient density profile counteracts this.

To estimate the duration, we adopt a fiducial index of $n=3/2$, giving
\begin{multline}
    t_{\mathrm{ss,end,gal}} \sim 9.75\times 10^{8}\,\mathrm{yr} \, \times \, \left( \frac{R_{w,0}}{100\,\mathrm{pc}} \right)^{-3} \times \\ \left( \frac{v_w}{{100\,\mathrm{km}\,\mathrm{s^{-1}}}} \right)^{-3}\times \\ \left( \frac{\dot{M}}{10^{-1}\,\mathrm{M}_{\odot}\,\mathrm{yr}^{-1}} \right)^{2}\times\left( \frac{n_{a,0}}{10^{-2}\,\mathrm{cm}^{-3}} \right)^{-2}\,,
\end{multline}
though this result is highly sensitive to the normalization of the CGM density profile, which is less constrained than $v_w$ and $\dot{M}$. Nevertheless, this estimate seems to indicate that the interaction phase is highly relevant for galactic winds.

\subsubsection{TDE outflows}

An $n=3/2$ scenario is also typical of the expected environments of some TDE outflows, given the natural stratification of accreting material around the central black hole \citep{1952MNRAS.112..195B} (though radio modelling often favours even steeper ambient profiles; e.g.~\citealt{2016ApJ...819L..25A, 2021ApJ...919..127C}). We use $R_{w,0} = 10^{13}\,\mathrm{cm}$, consistent with the simplified scenario presented in \cite{2009MNRAS.400.2070S} where the outflow launch radius is $\sim$ twice the pericenter distance of the stellar orbit around the black hole. Extrapolating typical densities of $n_a \sim 100\,\mathrm{cm}^{-3}$ at $\sim 0.01\,\mathrm{pc}$ \citep{2016ApJ...819L..25A} back to $10^{13}\,\mathrm{cm}$ for $n=3/2$ yields $n_{a,0} \sim 10^7\,\mathrm{cm}^{-3}$. For total disruptions, peak fallback accretion rates range from a few $\mathrm{M}_{\odot}$ to tens of $\mathrm{M}_{\odot}$ per year \citep{2024ApJ...961L...2B,2025ApJ...990..104F}, but not all of this material will be ejected in the outflow and the peak fallback rate is (clearly) greater than the average fallback rate. Partial disruptions can also yield substantially smaller fallback rates (e.g., \citealt{2013ApJ...767...25G, 2020ApJ...899...36M, 2026ApJ...998...81B}). Taking $\dot{M} = 10^{-1}\,\mathrm{M}_{\odot}\,\mathrm{yr}$ and $v_{\rm w} = 10^4\,\mathrm{km}\,\mathrm{s}^{-1}$, we estimate an initial overdensity $f \sim 10^5$, indicating these outflows are heavily interaction-dominated.

For the duration of the interaction phase, we then estimate
\begin{multline}
    t_{\mathrm{ss,end,TDE}} \sim 2.86 \times 10^7\,\mathrm{yr} \, \times \, \left( \frac{R_{w,0}}{10^{13}\,\mathrm{cm}} \right)^{-3} \left( \frac{v_w}{{10^4\,\mathrm{km}\,\mathrm{s^{-1}}}} \right)^{-3} \\ \times \left( \frac{\dot{M}}{10^{-1}\,\mathrm{M}_{\odot}\,\mathrm{yr}^{-1}} \right)^{2}\left( \frac{n_{a,0}}{10^7\,\mathrm{cm}^{-3}} \right)^{-2}\,.
\end{multline}
Of course, the accretion-stratified structure of the ambient medium would not extend far enough to support such a long interaction phase, so in reality these systems would be limited by the radial extent of the surrounding gas profile. \textbf{Furthermore, the outflow may shut off much earlier than $t_{\rm ss,end}$. If the wind is tied to the super-Eddington phase of the TDE accretion flow, then the duration of such a wind can be reasonably estimated to be $\sim$ months \citep{2011MNRAS.415..168S, 2018MNRAS.478.3016W}. Conservatively adopting a duration of 1 month, $t_{\rm w,e} \gg t_{\rm dyn}$ based on the same parameters used to estimate $t_{\rm ss,end}$. Thus, the time at which the reverse shock is expected to fully cross the finite-width ejecta shell is}
\begin{equation}
    t_{\Delta} \sim 8.24\,\mathrm{yr}\,\times \left( \frac{t_{\rm w,e}}{2.6\times 10^6\,\rm{s}} \right) \times \left( \frac{f}{10^5} \right)^{2/5}\, .
\end{equation}
\textbf{While this timescale is much shorter than $t_{\rm ss,end}$, radio observations of TDE outflows are carried out within $\sim$ months to years after the initial disruption. As such, these systems are expected to remain interaction-dominated throughout much of their observationally relevant evolution.}

\subsubsection{LFBOTs}

The analysis here may also be relevant to LFBOTs, whose radio and X-ray properties in at least some well-studied cases suggest fast outflows interacting with dense circumstellar material \citep[e.g.]{Margutti_2019, 2019ApJ...871...73H}. Recent modelling of the LFBOT AT2024wpp suggests that the circumstellar density structure approximately adheres to a power law where $n$ is as high as $\sim 3$ \citep{2025ApJ...993L...6N,2026arXiv260103337P}. Such steep power-laws are consistent with modelling assumptions for other similar events such as AT2018cow \citep{2019ApJ...871...73H}, where it is often assumed that the emission is coming from an outflow propagating through a wind-like ($n \sim 2$) circumstellar medium \citep{2022ApJ...935..157G,Margutti_2019}. These systems are therefore especially compelling from the standpoint of the present work, because they occupy precisely the regime in which the interaction phase is most prolonged. 

Indeed, for $n \geq 2$ the flow never relaxes to an energy-conserving similarity state; instead, the interaction-dominated structure persists indefinitely. In these scenarios, the ambient density declines more rapidly with radius than that of the freely expanding wind, and hence the reverse shock never propagates to small radii relative to the contact discontinuity. Therefore, if the circumstellar media for these events truly are wind-like or steeper, then these systems could remain in the interaction regime perpetually. Direct application of the interaction solutions to emission modelling in such transients is therefore a particularly natural next step, though outside the scope of this work. \textbf{As discussed in prior sections, the relevance of these interaction solutions could potentially be limited by the duration of the outflow. However, the properties of these systems are not very well-constrained, so it is difficult to assess the importance of this.}

\section{Summary and conclusions}
\label{sec:conclusions}

We presented a unified study of the early adiabatic evolution of wind-driven bubbles expanding into power-law ambient media, focusing on the relationship between the interaction-dominated early phase and the later energy-conserving phase, each of which has a self-similar solution. Using analytic scalings together with one-dimensional shock-capturing hydrodynamic simulations, we quantified both the timescales over which the flow relaxes toward these asymptotic states and the degree to which the corresponding solutions reproduce the full shocked structure. Our principal conclusions are:

\begin{enumerate}
    \item In the limit of highly overdense ejecta, the interaction solutions provide an excellent description of the shocked flow at early times. For all density profiles considered here, $0 \leq n \leq 2$, the numerical solutions rapidly relax toward the interaction-phase profiles when $f \gg 1$, where $f$ is the initial ratio of the wind density to the ambient density.
    \item For $n \leq 2$, both the analytic estimate derived in Section~\ref{sec:basic-considerations} and the numerical results in Section~\ref{sec:results} show that the duration of the interaction regime increases strongly with both $f$ and the ambient density power-law slope, following $t_{\mathrm{ss,end}} \propto f^{1/(2-n)}$. Physically, the interaction solutions cease to apply once the shocked-wind shell (i.e., the fluid bounded by the reverse shock and the contact discontinuity) is no longer thinner than the forward-shocked ambient shell, which in practice occurs when the instantaneous overdensity has fallen to $f(R_{\mathrm{c}}) \sim$ a few. At later times, the flow approaches the energy-conserving phase, with the shock positions converging toward the expected power-law scalings.
    \item The case $n=2$ is qualitatively distinct. In this limit, the predicted transition timescale diverges, and our simulations confirm that the system does not relax to a distinct energy-conserving similarity state. Rather, the interaction-dominated structure persists throughout the evolution, which is consistent with the fact that the ratio of the reverse shock radius to the contact discontinuity radius does not approach zero at late times.
    More generally, our results imply that steep ambient density gradients lead to prolonged interaction-dominated evolution and delay -- or even entirely prevent -- the onset of the later energy-conserving regime.
    \item The transition between these regimes is itself dynamically extended. Near $t \sim t_{\mathrm{ss,end}}$, neither the interaction nor the energy-conserving solution provides an accurate representation of the full shocked structure, especially in the velocity and pressure profiles. Even under a conservative error criterion, the interval over which the system is not particularly well described by either asymptotic solution can extend for several orders of magnitude in time beyond the nominal transition timescale itself. This indicates that it may generally be insufficient to model the evolution as an abrupt hand-off from one self-similar regime to the other, \textbf{and we developed a set of closed-form, analytic equations for the discontinuity positions as functions of time in Appendix~\ref{sec:analytic-functions}. To adequately account for the evolution of the full radial profiles, there is room for future work here in exploring the higher-order corrections to the interaction solutions.}
    \item These results have important implications for engine-driven transients. Although the interaction phase is shorter than the total non-radiative lifetime for classical O-star wind bubbles, it can still persist for decades. Galactic winds in stratified halos, TDE outflows, and at least some LFBOTs may remain interaction-dominated throughout much of their observed lifetimes.
\end{enumerate}

There are several natural extensions of this work. On the theoretical side, it would be useful to incorporate higher-order corrections to the interaction solutions and to develop a more quantitative description of the broad intermediate regime between the two asymptotic states. On the astrophysical side, coupling these hydrodynamic solutions to explicit radiation post-processing would allow more direct comparisons with radio, X-ray, and other observational diagnostics in systems such as TDEs and LFBOTs. Such extensions would help clarify when the interaction phase is merely an early hydrodynamic transient and when it is, in practice, the dominant regime governing the observable evolution of the source.

\section*{Acknowledgements}
\textbf{We thank the anonymous referee for their useful and constructive comments.} B.A. and E.R.C.~acknowledge support from NASA through the Astrophysics Theory Program, grant 80NSSC24K0897, and through Chandra Award Number 25700383 issued by the Chandra X-ray Observatory Center, which is operated by the Smithsonian Astrophysical Observatory for and on behalf of the National Aeronautics Space Administration under contract NAS8-03060.

\renewcommand{\thefigure}{B\arabic{figure}}
\setcounter{figure}{0}
\setcounter{equation}{0}
\renewcommand{\theequation}{A\arabic{equation}}

\begin{appendix}
\section{1D SRHD code}
\label{sec:srhd_code}
We solve the one-dimensional special relativistic hydrodynamics equations in spherical coordinates using a moving-mesh finite volume method. The equations are written in conservative form,
\begin{equation}
    \frac{\partial \mathbf{U}}{\partial t} + \frac{1}{r^2} \frac{\partial (r^2 \mathbf{F})}{\partial r} = \mathbf{S},
\end{equation}
where the source term $\mathbf{S}=(0, 2p/r, 0)^T$ accounts for geometric factors arising from spherical coordinates, following \citep{2006ApJS..164..255Z}, and $\mathbf{U} = (D, S, E)^T$ contains the lab-frame mass density $D = \Gamma\rho$, momentum density $S = \Gamma^2 \rho h u$, and energy density $E = \Gamma^2 \rho h - p - D$. Here, $\rho$ is the comoving mass density, $u = \Gamma\beta$ is the radial four-velocity, $p$ is the gas pressure, $\Gamma = \sqrt{1 + u^2}$ is the Lorentz factor, and $h = 1 + \hat{\gamma} p / [(\hat{\gamma}-1)\rho]$ is the specific enthalpy for an ideal gas with adiabatic index $\hat{\gamma}$. The corresponding flux vector is
\begin{equation}
    \mathbf{F} = \begin{pmatrix} Dv \\ Sv + p \\ Ev + pv \end{pmatrix},
\end{equation}
where $v = u/\Gamma$ is the three-velocity.

The spatial discretization uses piecewise linear reconstruction to achieve second-order accuracy in smooth regions. Slopes are limited using the minmod limiter,
\begin{equation}
    \sigma = \frac{1}{4}|\mathrm{sgn}(a) + \mathrm{sgn}(b)| \cdot
    (\mathrm{sgn}(a) + \mathrm{sgn}(c)) \cdot
    \min(|a|, |b|, |c|),
\end{equation}
where $a = \theta(q_i - q_{i-1})$, $b = (q_{i+1} - q_{i-1})/2$, $c = \theta(q_{i+1} - q_i)$, and $\theta = 1.5$ controls the limiter's compressiveness. This prevents spurious oscillations near discontinuities while maintaining accuracy in smooth flow regions.

Numerical fluxes at cell interfaces are computed using the HLLE approximate Riemann solver. Given left and right states at an interface, the relativistic wave speeds are estimated as
\begin{equation}
    a^{\pm}_{L,R} = \frac{v_{L,R} \pm c_{s}}{1 \pm v_{L,R} c_{s}},
\end{equation}
where $c_s = \sqrt{\hat{\gamma} p / \rho h}$ is the relativistic sound speed. The bounding wave speeds $a^- = \min(a^-_L, a^-_R)$ and $a^+ = \max(a^+_L, a^+_R)$ define the extent of the Riemann fan, and for a face moving with speed $v_F$, the ALE-HLLE numerical flux is given by
\begin{equation}
    \hat{\mathbf{F}} =
    \begin{cases}
        \mathbf{F}_L - v_F\mathbf{U}_L\,, & v_F \leq a^-\,,\\
        \mathbf{F}^* - v_F\mathbf{U}^*\,, & a^- < v_F < a^+\,,\\
        \mathbf{F}_R - v_F\mathbf{U}_R\,, & v_F \geq a_+\,,
    \end{cases}
\end{equation}
with
\begin{equation}
    \mathbf{F}^* = \frac{a^+ \mathbf{F}_L - a^- \mathbf{F}_R +
    a^+ a^- (\mathbf{U}_R - \mathbf{U}_L)}{a^+ - a^-}
\end{equation}
and
\begin{equation}
    \mathbf{U}^* = \frac{a^+\mathbf{U}_R - a^-\mathbf{U}_L - (\mathbf{F}_R - \mathbf{F}_L)}{a^+ - a^-}\,.
\end{equation}

Recovery of primitive variables $(\rho, u, p)$ from the conserved quantities $(D, S, E)$ requires solving a nonlinear equation for the pressure. We employ Newton-Raphson iteration with a convergence tolerance of $10^{-12}(D + E)$.

A key feature of our method is the moving-mesh formulation, wherein the computational domain is bounded by tracked discontinuities rather than fixed spatial boundaries. The global domain is divided into two subdomains, each consisting of $N$ uniformly-spaced grid cells. The outer edge of the outer subdomain coincides with the forward shock, while the inner edge of the inner subdomain tracks the reverse shock, and the boundary separating the two subdomains follows the contact discontinuity separating the shocked ejecta from the shocked ambient medium. Shock positions are evolved using speeds determined from the exact Rankine-Hugoniot jump conditions, given the upstream (unshocked) and downstream (shocked) states. The upstream state is supplied as a boundary condition representing either the unshocked ambient medium (for the forward shock) or the freely-expanding ejecta (for the reverse shock). At the contact discontinuity, the interface moves at the common fluid velocity shared by the adjacent zones.

Time integration is performed using a third-order strong stability-preserving Runge-Kutta (SSP-RK3) scheme:
\begin{align}
    \mathbf{U}^{(1)} &= \mathbf{U}^n + \Delta t \, \mathcal{L}(\mathbf{U}^n), \\
    \mathbf{U}^{(2)} &= \tfrac{3}{4} \mathbf{U}^n + \tfrac{1}{4}
        \mathbf{U}^{(1)} + \tfrac{1}{4} \Delta t \, \mathcal{L}(\mathbf{U}^{(1)}), \\
    \mathbf{U}^{n+1} &= \tfrac{1}{3} \mathbf{U}^n + \tfrac{2}{3}
        \mathbf{U}^{(2)} + \tfrac{2}{3} \Delta t \, \mathcal{L}(\mathbf{U}^{(2)}),
\end{align}
where $\mathcal{L}$ denotes the spatial discretization operator. The timestep is determined by a CFL condition, $\Delta t = C_{\rm cfl} \, \Delta r_{\rm min} / \lambda_{\rm max}$, where $\lambda_{\rm max}$ is the maximum signal speed and $C_{\rm cfl} = 0.3$.

\section{Numerical convergence of simulations}
\label{sec:numerical-convergence}

\begin{figure*}
\includegraphics{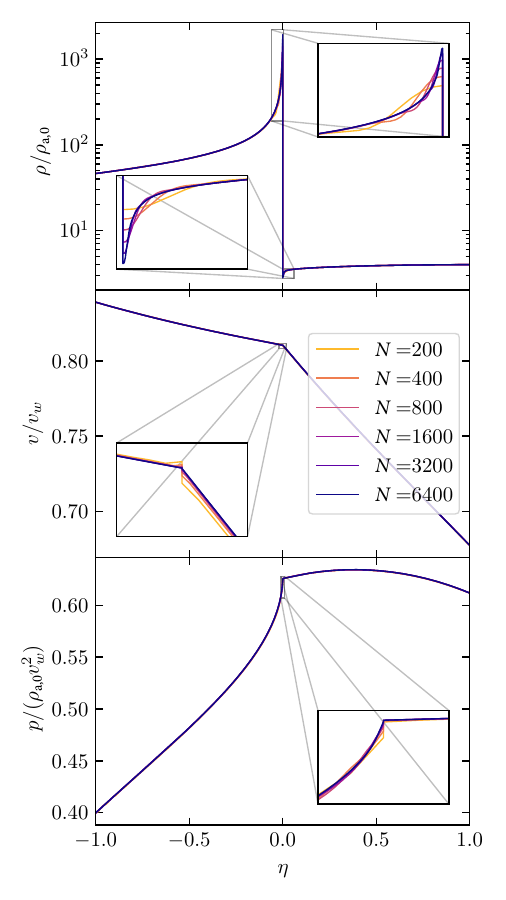}
\includegraphics{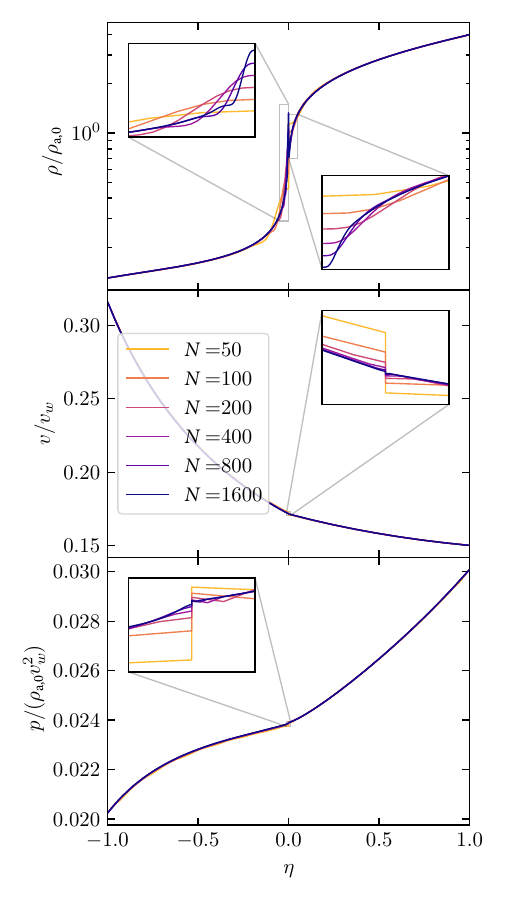}
\caption{A numerical convergence test for the profiles of the fluid mass density, velocity, and pressure vs. self-similar coordinate $\eta$ at six different resolutions. This simulation was initialized with a wind velocity $v_w=0.01$ and an overdensity $f=10^3$, impacting a uniform ($n=0$) ambient medium. These snapshots were taken at $t/t_{\mathrm{dyn}}=10$ (\textbf{left}) and $t/t_{\mathrm{dyn}}=10^6$ (\textbf{right}) to explore convergence in both the interaction and energy-conserving regimes. Broadly, the profiles are extremely well-converged with resolutions even as low as $N=200$ zones per subdomain, though the immediate vicinity of the contact (within $\sim 10$ zones) remains under-resolved even for large $N$.}
\label{fig:resolution_tests}
\end{figure*}

The shock-capturing method outlined in Appendix~\ref{sec:srhd_code} allows for a high degree of numerical accuracy with relatively few zones. All fluid profiles are generally very well converged even for resolutions as low as $N=200$ zones per subdomain, though the behaviour near the contact is extremely challenging to resolve. Because this study is primarily concerned with broader agreement with existing self-similar solutions and associated transitional timescales, as well as shock dynamics specifically, we adopt a baseline resolution of $N=800$ while acknowledging that the density cusps adjacent to the contact are slightly under-resolved in both the interaction and energy-conserving regimes. The under-resolved features near the contact can be seen in the insets in Figure~\ref{fig:resolution_tests}; outside of this narrow region, however, the simulated profiles are broadly numerically converged even for lower $N$.

\renewcommand{\thefigure}{C\arabic{figure}}
\setcounter{figure}{0}
\setcounter{equation}{0}
\renewcommand{\theequation}{C\arabic{equation}}

\section{Analytic functions for discontinuity positions}
\label{sec:analytic-functions}

\textbf{The time-dependent positions of all three relevant discontinuities - the forward shock, contact discontinuity, and reverse shock - have analytic expressions in both the interaction and energy-conserving limits. We can smoothly connect these expressions to produce a set of formulae for these positions that can be explicitly evaluated at any time $t$, given the necessary parameters. We adopt the smooth interpolation function}
\begin{equation}
    S(t) = \frac{t^{2p}}{t^{2p} + qt_{\rm ss,end}^{2p}}\,,
\end{equation}
\textbf{where $t_{\mathrm{ss,end}}$ is the transition timescale in Equation~\ref{eq:tss}, $q=1.11-0.64n$, and $p=1.16-0.53n$ (these latter two values were obtained via fitting). Using this function to conjoin the analytic expressions in each limiting regime, we obtain}
\begin{multline}
    R_{\rm s}(t) = [1-S(t)](R_0 + V_{\rm ej}t)(1+\Delta_{\rm s})\\
    \times \left[ 1 + \left( \frac{\kappa_{\rm s}}{1 + \Delta_{\rm s}} - \kappa_{\rm c} \right)\frac{1}{\sqrt{f}} \left( \frac{R_0 + V_{\rm ej}t}{R_0} \right)^{(2-n)/2} \right]\\
    + S(t)K\left( 2\pi R_0^{2-n}V_{\rm ej}^3 f \right)^{1/(5-n)}t^{3/(5-n)}
    \label{eq:forward-shock-model}
\end{multline}
\begin{multline}
    R_{\rm c}(t) = [1-S(t)](R_0 + V_{\rm ej}t)\\
    \times \left[ 1 - \frac{\kappa_{\rm c}}{\sqrt{f}} \left( \frac{R_0 + V_{\rm ej}t}{R_0} \right)^{(2-n)/2} \right]\\
    + S(t)\xi_{\rm c}K\left( 2\pi R_0^{2-n}V_{\rm ej}^3 f \right)^{1/(5-n)}t^{3/(5-n)}
    \label{eq:contact-model}
\end{multline}
\begin{multline}
    R_{\rm r}(t) = [1-S(t)](R_0 + V_{\rm ej}t)\\
    \times \left[ 1 - \frac{(\kappa_{\rm c} + \kappa_{\rm r})}{\sqrt{f}} \left( \frac{R_0 + V_{\rm ej}t}{R_0} \right)^{(2-n)/2} \right]\\
    + S(t)\sqrt{\frac{\gamma+1}{\gamma-1}\left( 1 - \frac{n+4}{9\gamma} \right)\frac{3}{(5-n)V_{\rm ej}}}\\ \times(\xi_{\rm c}K)^{3/2}\left( 2\pi R_0^{2-n}V_{\rm ej}^3 f \right)^{3/(10-2n)}t^{(n+4)/(10-2n)}\,.
    \label{eq:reverse-shock-model}
\end{multline}

\textbf{Here, $f_0$, $R_0$, and $V_{\mathrm{ej}}$ are initial conditions. All other parameters are extracted from either the self-similar solutions presented in \cite{2024ApJ...975L..14C} or by numerically integrating to obtain the contact position in the energy-conserving limit. Values for these parameters are tabulated in Table~\ref{tab:parameters} for several different values of $n$ and a fixed value of $\gamma=5/3$. Note that $n=2$ is not included here since the system will remain in the interaction-dominated state indefinitely, and would therefore be described solely by the solutions presented in \cite{2024ApJ...975L..14C}. To further minimize the error in the transition window for the full radial profiles for the density, velocity, and pressure, higher-order terms can be included in these same interaction solutions.}

\setcounter{table}{0}
\renewcommand{\thetable}{C\arabic{table}}

\begin{table}[b!]
 \caption{Self-Similarity Parameters ($\gamma=5/3$)}\label{sample-table}
 {\tablefont\begin{tabular}{@{\extracolsep{\fill}}lcrrrrr}
   \toprule
    $n$ & $K$ & $\xi_c$ & $\Delta_{\rm s} \times 10$ & $\kappa_{\rm s} \times 10$ & $\kappa_{\rm c} \times 10$ & $\kappa_{\rm r} \times 10$ \\
    \hline
    0.00 & 0.8837 & 0.8584 & 0.9572 & 0.3350 & 3.9526 & 0.4852 \\
    0.25 & 0.8531 & 0.8557 & 1.0199 & 0.3405 & 4.2295 & 0.5263 \\
    0.50 & 0.8184 & 0.8531 & 1.0914 & 0.3431 & 4.5495 & 0.5751 \\
    0.75 & 0.7790 & 0.8505 & 1.1740 & 0.3412 & 4.9240 & 0.6341 \\
    1.00 & 0.7340 & 0.8480 & 1.2703 & 0.3314 & 5.3691 & 0.7069 \\
    1.25 & 0.6826 & 0.8457 & 1.3843 & 0.3084 & 5.9078 & 0.7990 \\
    1.50 & 0.6238 & 0.8435 & 1.5214 & 0.2621 & 6.5756 & 0.9196 \\
    1.75 & 0.5566 & 0.8416 & 1.6898 & 0.1731 & 7.4298 & 1.0845
    \botrule
    \label{tab:parameters}
    \end{tabular}}
\end{table}

\begin{figure}
\includegraphics{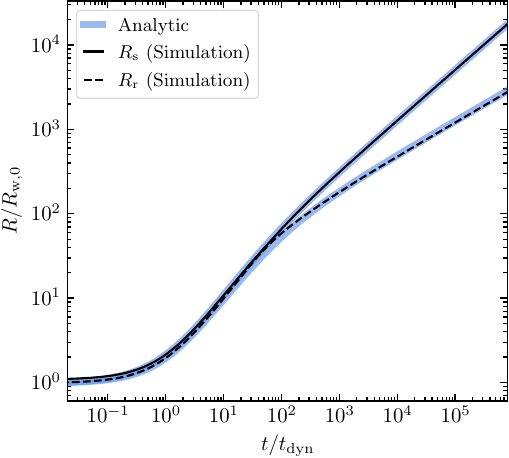}
\caption{\textbf{Comparison of simulated forward and reverse shock positions to the analytic functions presented in Equations~\ref{eq:forward-shock-model}-\ref{eq:reverse-shock-model} for $n=0$ and $f=10^3$.}}
\label{fig:analytic-comparison-1}
\end{figure}

\begin{figure}
\includegraphics{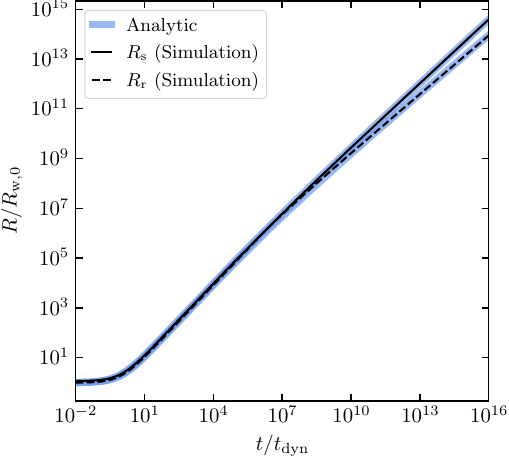}
\caption{\textbf{Comparison of simulated forward and reverse shock positions to the analytic functions presented in Equations~\ref{eq:forward-shock-model}-\ref{eq:reverse-shock-model} for $n=1.5$ and $f=10^3$.}}
\label{fig:analytic-comparison-2}
\end{figure}

\textbf{A direct comparison between these equations and the simulation output is shown in Figures~\ref{fig:analytic-comparison-1}\&\ref{fig:analytic-comparison-2} for $f=10^3$ and $n=0,1.5$.}

\end{appendix}

\end{document}